# The Well-Being Palette: An Action-Word Selection Tool Designed for Low-Burden Reflection on Workplace Well-Being

**Nobuhiko Muramoto[1], Takayuki Nagaya[1], Tomoko Tanaka[1], Koichiro Iwai[1], Katsunori Kohda[1]**

[1]Toyota Central R&D Labs., Inc., Nagakute, Japan

**Abstract**

**Background:** Worker well-being is important for sustainable productivity, organizational health, and quality of work life. However, many workplace well-being interventions require time or effort, making them difficult to integrate into daily routines. **Objective:** This study developed and explored a Well-Being Palette. This tool asks workers to select action-oriented well-being words when reflecting on positive experiences during the workday. **Methods:** We conducted a three-month exploratory field study in a corporate workplace. Ninety participants took part in the study, and 88 were included in the primary analysis after excluding two participants with missing PERMA data. Participants used the application from October 1 to December 26, 2024. Subjective well-being was assessed using the PERMA Profiler at baseline, one month after the start, two months after the start, immediately after the intervention (three months), and at a one-month follow-up. Application logs were analyzed to identify usage patterns, departmental patterns in word selection, and selected-word diversity using Shannon entropy. A subset of participants also took part in well-being workshops. **Results:** During the formal intervention period, participants generated 3,480 input records containing 10,104 selected action words and selected all 72 available words. Overall PERMA scores increased from $5.90 \pm 1.30$ at baseline to $6.28 \pm 1.33$ immediately after the intervention ($t[87] = 3.65$, $p < .001$, Cohen's $dz = 0.39$) and remained similar at the one-month follow-up among available cases. Departmental patterns in the selected word categories suggested differences in how well-being-related experiences were recognized across contexts. Workshop participants showed a more sustained upward PERMA pattern and a larger cumulative increase in selected-word diversity quantified using Shannon entropy. Among the workshop participants, lower pre-workshop alpha diversity was associated with larger subsequent increases, suggesting that social reflection may be relevant for participants with narrower selected-word repertoires. **Conclusion:** These findings suggest that brief action-word reflection can be integrated into workplace routines and may support the recognition of well-being-related experiences. Shannon entropy may provide a complementary lens for characterizing the diversity of selected reflective words. However, the findings should not be interpreted as causal evidence of objective changes outside the application.

## 1 Introduction

Worker well-being is recognized as a central issue in sustainable productivity, organizational health, and quality of work life. Positive psychology has contributed to this shift by emphasizing both the reduction of ill-being and the cultivation of well-being and human flourishing (Seligman, 2019). In organizational settings, well-being is closely linked to work outcomes and broader functioning, making it a practical and psychological concern (Harter et al., 2003; Diener et al., 2015).

As workers spend most of their waking time engaged in work-related activities, everyday workplace experiences may play a central role in shaping their well-being (Kahneman et al., 2004). Positive psychological interventions have shown that relatively simple reflective practices, such as noticing positive experiences or expressing gratitude, can support subjective well-being (Emmons and McCullough, 2003; Seligman et al., 2005). Meta-analytic evidence further suggests that positive psychological interventions can improve well-being and related mental health outcomes across diverse formats and populations, although effect sizes and durability vary by intervention type and study context (Hendriks et al., 2020; Carr et al., 2021).

Digital technologies offer scalable methods for delivering such interventions in daily life. Mobile- and web-based interventions have shown promise in improving mental health and well-being; however, engagement and sustained adherence are persistent challenges (Linardon et al., 2019; Groot et al., 2023). Many digital interventions require structured exercises, repeated task completion, or additional time commitments, increasing the cognitive or temporal burden and reducing sustained use (Diefenbach, 2018; Baumel et al., 2019). These issues are particularly important in workplaces where time, attention, and mental resources are constrained. Accordingly, there is a need for low-burden, user-centered tools that can be integrated into existing work routines, while still supporting reflection (Mohr et al., 2014; Schueller and Torous, 2020).

In this study, we developed the Well-Being Palette, a web-based reflection tool designed to support brief daily reflections on well-being-related actions and experiences. The tool presents 72 action-oriented well-being words in a mandala-style grid-based structure. At the end of each workday, users briefly review the structure and select up to three words that best represent the positive actions or experiences they encountered when reflecting on that day. This approach is action-based, in that well-being is presented through words that users can connect to daily experiences, rather than as an abstract psychological state.

Conceptually, this action-word selection approach differs from open-ended journaling, gratitude exercises, or mood tracking by providing a structured and brief way for users to notice and label concrete well-being-related experiences. However, little is known about whether the distribution of selected reflective words can be used to characterize the breadth of individuals' well-being-related repertoires beyond changes in average well-being scores.

Accordingly, this study also examined whether application log data could provide a complementary perspective beyond conventional score-based assessments. Most well-being studies have focused on the level of subjective well-being, such as mean changes in questionnaire scores. However, two individuals with similar well-being scores may differ in the range of experiences or action words associated with their well-being. To characterize this distributional aspect, we applied Shannon entropy to the distribution of the selected words. In the present study, selected-word diversity refers to the diversity of selected well-being-related action words, reflecting the breadth of the participants' reflective well-being repertoire.

We also explored whether structured sharing workshops could support social reflection. Here, the participants discussed their selected words and reasons for their choices, reviewed group-level selection patterns, and reflected on what they had learned. This design was based on the idea that exposure to others' interpretations of well-being-related experiences might broaden how individuals recognize and describe their own well-being.

We conducted a three-month exploratory field study in a corporate workplace using the Well-Being Palette application. We examined application usage, changes in subjective well-being measured using PERMA Profiler, selected-word diversity quantified using Shannon entropy, departmental patterns in word selection, and exploratory differences associated with participation in the sharing workshop. This study makes three notable contributions to the literature. First, it introduces an action-based reflection tool designed to be low-burden for workplace well-being. Second, it reports an exploratory field study that combines a questionnaire-based well-being assessment with daily application logs in a real-world workplace setting. Third, it proposes selected-word diversity, quantified using Shannon entropy, as a complementary lens for understanding how workers recognize and reflect on well-being-related actions and experiences.

## 2 Materials and Methods

### 2.1 Study Design

This study used a longitudinal exploratory field design to examine an action-based well-being intervention in workplace settings. The formal intervention period was from October 1 to December 26, 2024. Participants used a web-based application over a three-month period, and selection patterns and changes in subjective well-being over time were observed. This study was designed to reflect real-world usage conditions, rather than controlled laboratory settings.

### 2.2 Participants

Ninety participants were recruited across multiple departments of a private-sector corporate research institute in Japan. The participants were full-time workers engaged in diverse professional roles, including research and administrative roles. Before individual employees were invited, the research team consulted managers of candidate departments and obtained departmental cooperation to allow members of those departments to receive information about the study. Members of cooperating departments were then invited to attend an explanatory session conducted by the research team. At the session, participants were informed that participation was voluntary, that choosing not to participate would not result in any disadvantage in the workplace, and that they could decide individually whether to take part. Written informed consent was obtained from all participants before participation. Two participants were excluded from the primary analysis because their PERMA Profiler responses had missing values that prevented the calculation of relevant scores. The final sample comprised 88 participants, made up of 55 men and 33 women. The age groups were as follows: nine participants in their 20s, 17 in their 30s, 21 in their 40s, 29 in their 50s, and 12 in their 60s. The sample included 53 administrators and 35 researchers. Forty-four participants attended the sharing workshop, and 44 did not. The participant characteristics are summarized in Table 1.

### 2.3 System Design

We developed a web-based application using the Well-Being Palette framework (Figure 1). The main input screen presented 72 action-oriented well-being words in a mandala-style, grid-based, reflective structure organized into eight categories of Emotional Well-Being. The users were asked to briefly

**Table 1.** Demographics, departmental, role, and workshop-participation characteristics of the 88 analyzed participants.

| Characteristic | Category | n | % |
|---|---|---|---|
| Total | Participants | 88 | 100.0 |
| Department | Dept A | 6 | 6.8 |
| Department | Dept B | 5 | 5.7 |
| Department | Dept C | 7 | 8.0 |
| Department | Dept D | 10 | 11.4 |
| Department | Dept E | 25 | 28.4 |
| Department | Dept F | 28 | 31.8 |
| Department | Dept G | 7 | 8.0 |
| Workshop participation | Workshop | 44 | 50.0 |
| Workshop participation | Non-workshop | 44 | 50.0 |
| Age group | 20s | 9 | 10.2 |
| Age group | 30s | 17 | 19.3 |
| Age group | 40s | 21 | 23.9 |
| Age group | 50s | 29 | 33.0 |
| Age group | 60s | 12 | 13.6 |
| Gender | Male | 55 | 62.5 |
| Gender | Female | 33 | 37.5 |
| Role | Administrative member | 53 | 60.2 |
| Role | Researcher | 35 | 39.8 |

Note. Department labels were anonymized.

review the structure at the end of each workday and select up to three words that best represented the positive actions or experiences they recognized when reflecting on that day. The selected words were displayed in the input area before submission so that the users could confirm their choices. This interaction was designed to be completed in a few minutes, minimizing the cognitive and temporal burdens.

To support daily use, participants received a reminder email near the regular end of the workday at 17:30. As the application emphasized an overview of the 72-word palette and related visualizations, participants were instructed to use it primarily on a desktop or laptop computer rather than on a mobile device. Individual selected-word records were accessible only to the participant and study administrators; other participants, including workplace supervisors, could not access individual-level entries through the application. Before the sharing workshops, participants could review their own selection data and aggregated group-level data, but they could not view other participants' individual records.

The application also included dashboard functions that allowed users to review their previous selections and recorded daily words. On the selected-day review screen, the previously selected words were marked on the palette with crown icons indicating the selection order: gold for the first selection, silver for the second, and bronze for the third. The system also visualized the accumulated records using word-by-date heat maps, weekly category-composition charts, selected-word frequency rankings by selection order, selection-frequency and co-occurrence maps, and circular co-occurrence visualizations. The daily input interface is shown in Figure 1, and the dashboard visualization functions are shown in

1st：良いところを褒める 2nd：成功を信じる 3rd：好きな仕事に貢献できる send reset

| 成長を感じる<br>Feel the growth | 笑顔で接する<br>With a smile | 相違するお互いを認める<br>Diversity & Inclusion | 環境の変化で生き残る<br>Survive | 多様化する<br>Diversify | 新しい結合を持つ<br>Innovate | 伝えたいことを持つ<br>Have something to tell | 移動する<br>Move | 競争する<br>Compete |
|---|---|---|---|---|---|---|---|---|
| 眼を見て話す<br>Look in eyes | 認める<br>Acknowledge | 握手する<br>Shake hands | 複数の定常状態を持つ<br>Multiple steady states | 進化する<br>Evolve | DNAを変える<br>Mutate | 体験する<br>Experience | 遊ぶ<br>Play | 新しいものに触れる<br>Associate with innovation |
| 貢献する<br>Contribute | 感謝する<br>Appreciate | 良いところを褒める<br>Praise | 価値観を変える<br>Change values | 絶滅する<br>Extinct | 新しい機能を創り出す<br>Functionalize | 脳内報酬系を働かす<br>Activate brain reward system | 運動する<br>Exercise | 熟練を実感する<br>Feel the progress |
| 記憶する<br>Remember | チームで考える<br>Think as a team | 偶然を楽しむ<br>Serendipity | 認める<br>Acknowledge | 進化する<br>Evolve | 遊ぶ<br>Play | 成長を期待する<br>Expect the growth | 一歩を踏み出す<br>Go forward | 成功を信じる<br>Believe in success |
| 思考する<br>Think deeply | 創造する<br>Create | アートする<br>Art | 創造する<br>Create | Emotional<br>Well-Being | 希望を持つ<br>Hope | 楽しむ<br>Enjoy | 希望を持つ<br>Hope | 幸せな未来を想う<br>Peaceful future |
| 見える化する<br>Visualize | 演じる<br>Play a role | 新しい関係性を構築する<br>Connect | 共生する<br>Symbiosis | 健康を感じる<br>Wellness | 触れ合う<br>Contact | 自身を知る<br>Understand myself | 困難に立ち向かう<br>Face a challenge | 能力を獲得する<br>Habilitate abilities |
| 多様性を活かし共創できる<br>Diversity and pluralism | 植物と触れ合う<br>Touch plants | 永続的な環境を持つ<br>Persistent environment | 匂いを感じる<br>Smell natural | おいしくご飯を食べる<br>Feel delicious | 好きな仕事に貢献できる<br>Work with love | 会話する<br>Communicate | 共感する<br>Empathy | 他者との繋がりを作る<br>Connect with others |
| 神仏を敬う<br>Revere nature | 共生する<br>Symbiosis | 困っている人を助けたいと思う<br>Help people in need | 家族を大切にする<br>Love family | 健康を感じる<br>Wellness | 信頼できる友人を持つ<br>Have great friends | 気配りする<br>Pay attention | 触れ合う<br>Contact | 身振りで伝える<br>Gesture |
| 平和を感じる<br>Feel the peace | 動物・昆虫と触れ合う<br>Touch creatures | 空気をおいしく感じる<br>Feel the air good | 病気から回復する<br>Recover from illness | 質の高い睡眠をとる<br>Have quality sleep | 自己決定する<br>Self-determine | 存在を実感する<br>Feel presence | お互いに教え合う<br>Teach each other | 伝染する<br>Contagion |

**Figure 1.** Daily interface presenting 72 action-oriented well-being words in eight categories around emotional well-being. Users selected up to three words after reflecting on positive actions or experiences from the workday.

Supplementary Figure S1. These visualizations were intended to support self-reflection and provide materials for the sharing workshop in which participants explained their own selection patterns to others.

## 2.4 Development of the Action-Word Set

The well-being action-word set was developed through an exploratory, practice-oriented process by five members of an internal well-being research group within the company. The group placed Emotional Well-Being at the center of the framework. It defined eight core categories based on workplace-relevant experiences of well-being: acknowledge, evolve, play, create, hope, symbiosis, wellness, and contact. Each category contained nine action-word labels, including a category anchor word and related action-oriented prompts, yielding a set of 72 words. These English category labels were used for presentation and interpretation, while individual words were treated as action-oriented reflection prompts.

The action-word set was developed as a practice-oriented vocabulary for reflection rather than as a direct adaptation of an existing psychometric taxonomy. To situate the word map conceptually, we referred to several strands of positive psychology and well-being research, including multidimensional models of well-being and positive functioning, character strengths, positive emotions, positive lexicography, and action-oriented well-being practices (Peterson and Seligman, 2004; Fredrickson, 2004; Huppert and So, 2013; Rusk and Waters, 2015; Lomas, 2016). These literatures collectively emphasize positive emotions, strengths, relationships, meaning, growth, vitality, awareness, and future-oriented action, but no single framework directly matched the intended brief word-selection format. The 72 words were therefore organized into eight broad categories to provide a balanced set of interpersonal, developmental, playful, creative, future-oriented, ecological, health-related, and embodied or relational prompts for daily reflection.

Words were expressed as verbs or action-oriented phrases to help users reflect on well-being as something enacted through daily behavior rather than as a passive state. English labels were prepared for the categories and words to support the presentation of the framework in English. These labels were treated as practical translations of the Japanese reflection vocabulary rather than as independently validated constructs. The number and structure of the words were determined based on visual clarity and usability within a grid-based, multidimensional reflective structure. Before implementation, the application interface and word-selection procedure were informally tested by a small number of internal users to check basic usability, readability, and whether the selection task could be completed as intended. This preliminary user testing was conducted as a practical implementation check rather than as a formal expert review, cognitive interview, or psychometric validation of the word set.

The word set should be interpreted as a configurable vocabulary for reflection, rather than a validated psychometric scale. Its purpose is to support low-burden self-reflection and word selection, and not to provide a fixed universal taxonomy of well-being.

## 2.5 Procedure

The study was conducted over three months, from October 1 to December 26, 2024. The participants were instructed to use the application daily after work. During each session, participants selected up to three words that represented experiences they perceived as positive when reflecting on that day. The follow-up assessment was conducted approximately one month after the end of the intervention period, with the final follow-up deadline on January 31, 2025. The overall study timeline, measurement points, and alpha diversity windows are shown in Figure 2.

In addition, a subset of participants took part in workshop-style sharing sessions. Here, they discussed their selected words and related their experiences with others. The workshops were conducted between November 13 and November 28, 2024, after the one-month assessment and before the two-month assessment.

## 2.6 Measures

### 2.6.1 Application Log Data

Application log data were obtained from a web-based system. Each record included an anonymized participant identifier, a timestamp, and up to three selected action-words. Timestamps were recorded at the date-time level and aggregated into date and monthly windows for analysis. The application log data included the number of input records, the number of input days per participant, the number of selected words per input, the frequency of selected words and categories, departmental selection patterns, and selected-word diversity. These records represented participants' recalled and self-recognized positive actions or experiences, as expressed through word selection; however, they did not provide objective verification that the corresponding experiences or actions occurred.

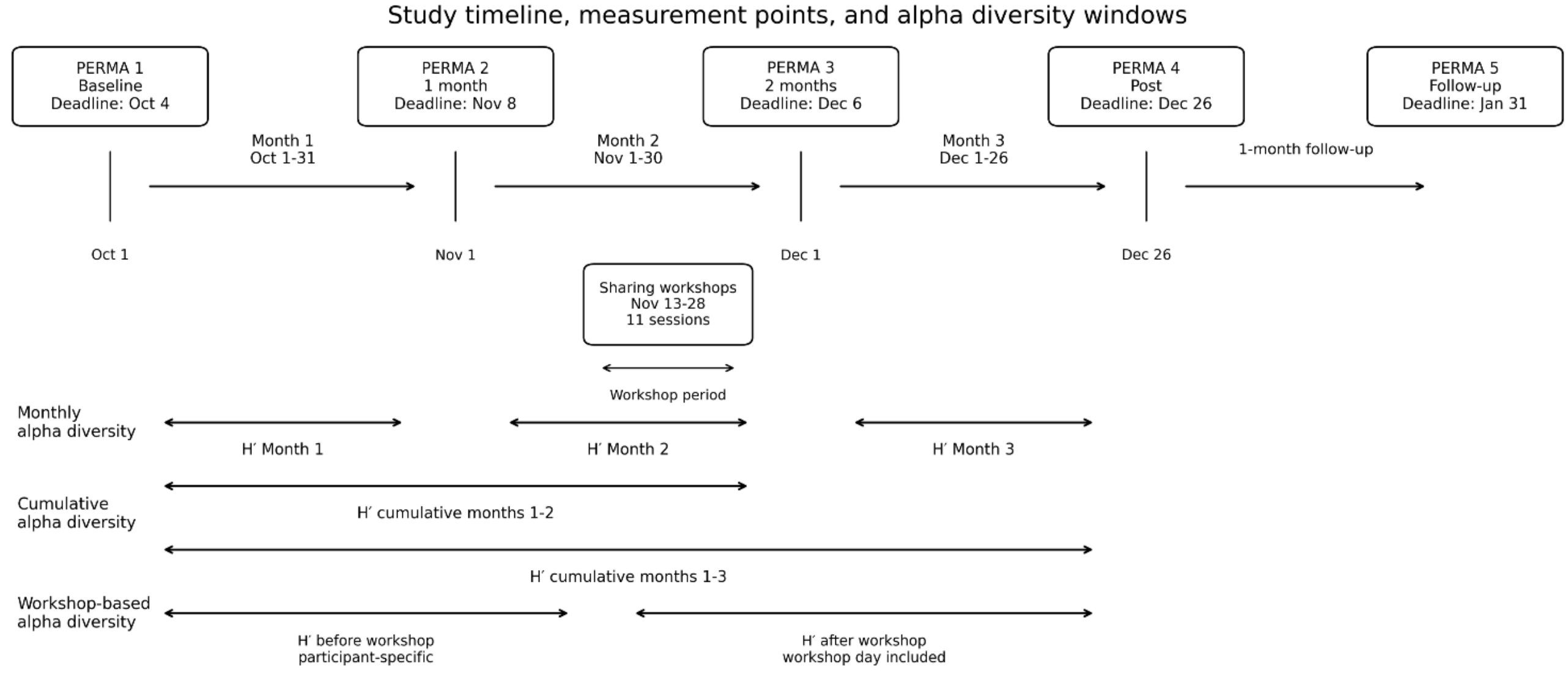

**Figure 2.** Timeline of PERMA Profiler assessments, monthly and cumulative alpha-diversity windows, and participant-specific workshop-based windows during the three-month field study.

### 2.6.2 Subjective Well-Being

Subjective well-being was assessed using the 23-item PERMA Profiler described by Butler and Kern (2016). A validated Japanese Workplace PERMA-Profiler is available for assessing well-being at work among Japanese workers (Watanabe et al., 2018). However, we used the general PERMA-Profiler structure, as the present study focused more broadly on subjective well-being and well-being-related experiences than on workplace-specific well-being alone.

A research team translated the questionnaire items and response anchors into Japanese for use with Japanese-speaking employees. The item structure, codes, domain assignments, and 0-10 response formats followed the PERMA Profiler. The original 23-item questionnaire includes the five core PERMA domains, negative emotion items, physical health items, a loneliness item, and a happiness item. However, only the five core PERMA domains and the overall PERMA score were analyzed in the present study.

The translated wording was reviewed internally to preserve the conceptual meaning of each item and ensure comprehensibility in the workplace context. The Japanese item wording and response anchors are provided in Supplementary Table S2. As this internal Japanese version was not independently validated or formally back-translated, PERMA scores were used in the present study as an exploratory index of subjective well-being rather than as evidence from a validated Japanese psychometric adaptation.

Each core PERMA domain score was calculated as the mean of three corresponding items. If one item within a three-item PERMA domain was missing, the domain score was calculated as the mean of the two completed items; domain scores were treated as missing if fewer than two items were available. The overall PERMA score was calculated as the mean of the 16 contributing items, consisting of the 15 items from the five PERMA domains and the single happiness item. When one or two contributing items were missing but all five domains could be scored and the happiness item was available, the overall score was calculated using the available contributing items. No statistical imputation was applied. In the analytic sample, baseline item-level responses included limited missingness: two participants had one missing contributing PERMA item, and one participant had two missing contributing PERMA items and one missing health item. All other item-level responses from baseline through post-intervention were complete.

Assessments were conducted at five time points: baseline (pre-intervention), one month after the start, two months after the start, immediately after the intervention (three months), and at a one-month follow-up after the intervention. PERMA Profiler data were available for all 88 participants at baseline, one month after the start, two months after the start, and immediately after the intervention. At the one-month follow-up, data were available for 74 participants. Therefore, analyses involving follow-up assessments were conducted using available cases.

#### 2.6.3 Selected-Word Diversity

Alpha diversity is an entropy-based diversity index that represents the variety of selected words. Let $c_i$ denote the number of times word $i$ was selected by a participant within a given time window, and let $N = \Sigma_i c_i$ denote the total number of selected words in that window. Shannon entropy was calculated as follows:

$$\boldsymbol{H'} = -\sum_{\boldsymbol{i=1}}^{\boldsymbol{72}} (\boldsymbol{p_i} \ln \boldsymbol{p_i})$$

where $p_i = c_i / N$. The natural logarithm was used, and entropy is therefore expressed in nats. Words with zero selections had $p_i = 0$ and contributed 0 to the sum; no pseudocounts were added. Entropy was calculated from proportions derived from raw selection counts. Because the same 72-word set was used for all participants and time windows, the primary analyses used unnormalized Shannon entropy. Normalized entropy, $H'/\ln(72)$, would rescale values to a 0-1 range but would not by itself address differences in the number of input opportunities across time windows. For each participant, alpha diversity was calculated based on the distribution of selected words within each target period. We calculated the monthly alpha diversity for October 1-31, November 1-30, and December 1-26. We also calculated the cumulative alpha diversity for the first two months and for the entire three-month intervention period.

For workshop participants, additional workshop-based alpha diversity indices were calculated by dividing the intervention period into two participant-specific windows. These were from the start of application use to the day before each participant's workshop and from the day of that participant's workshop to the end of the intervention. The workshop day was included in the post-workshop window because the workshop itself was expected to affect word selection.

Shannon entropy is widely used in ecological and microbiome studies (Shannon, 1948; Magurran, 2004), but has also been applied to human activity patterns and urban diversity analysis (Song et al., 2010; Yoshimura et al., 2022). To our knowledge, its application to well-being-related reflection data remains limited; therefore, in this study, we extended entropy-based diversity to capture the distribution of selected well-being-related words rather than their intensity alone.

## 2.7 Analysis

To examine the effects of the intervention, we conducted exploratory analyses. These included a pre-post comparison between baseline and immediately after the intervention, a follow-up comparison between immediately after the intervention and the 1-month follow-up, an analysis of differences across departmental groups, and comparisons between participants with and without sharing sessions.

Descriptive statistics were calculated for the application usage, word selection, PERMA scores, and alpha diversity. PERMA scores and alpha diversity values were reported as means and standard deviations. Changes in PERMA scores from baseline to immediately after the intervention and from immediately after the intervention to the one-month follow-up were examined using paired-sample t-tests as primary analyses. Wilcoxon signed-rank tests were also conducted as sensitivity analyses to confirm that the results were robust to the distributional assumptions. Differences between workshop and non-workshop participants in score levels and change scores were examined using Welch's t-test, with Mann-Whitney U tests used for non-parametric sensitivity analyses.

Changes in alpha diversity were examined separately for the month-, cumulative-, and workshop-based windows. For the workshop participants, the association between pre-workshop alpha diversity and subsequent changes was examined using Pearson's and Spearman's correlation coefficients. The effect sizes were calculated using Cohen's dz for paired comparisons and Cohen's d for group comparisons. As the analyses were exploratory, the results were interpreted based on the pattern of estimates, effect sizes, and p-values rather than on confirmatory hypothesis testing alone. Missing data were handled on an analysis-by-analysis basis. The number of participants included in each analysis is reported in the Results section.

Given the exploratory nature of the field study, statistical tests were used to characterize patterns in the data rather than to support confirmatory causal inference. The baseline-to-post-intervention change in the overall PERMA score was treated as the main outcome of interest. Domain-level PERMA analyses, follow-up comparisons, workshop-related analyses, selected-word diversity analyses, departmental comparisons, and correlation analyses were considered exploratory. P values were reported without formal adjustment for multiple comparisons; therefore, findings from exploratory analyses should be interpreted as hypothesis-generating, with emphasis placed on effect sizes and the consistency of observed patterns.

## 2.8 Well-Being Sharing Workshop

Sharing of well-being was conducted in a workshop format. Eleven workshop sessions were held between November 13 and 28, 2024, with four participants per session. Each session consisted of members from the same department so that participants could understand each other's work without an extensive explanation. Workshop participants were selected and allocated to groups to balance age, sex, and managerial status; participation was not based on the number of application inputs. Each workshop lasted approximately 75-90 minutes.

First, each participant shared their individual inputs, including the selected words and reasons for their choices. Then, the other participants asked questions to deepen their understanding. This process was repeated for all the participants. Second, the aggregated group-level data, including frequently selected words, are presented. Third, the participants conducted structured retrospective reflection using a modified "YWT" format, a Japanese retrospective framework similar in function to the keep-problem-try-style reflection. The three prompts are interpreted as Y: what was done, W: what was learned, and T: what to try next. In the present workshop, we adapted the first prompt to Y: "What went well?" because all participants had completed the same application-use task. The purpose was to elicit positive experiences and insights rather than to discuss completed actions. The W and T remained the same. Participants wrote short free-text responses on sticky notes for each of these prompts and organized them on a shared board. Finally, participants also wrote sticky-note responses to the prompt, "What does the Well-Being Palette mean to you?"

The workshop was treated as a multi-component social reflection session. The exploratory associations between PERMA scores and selected-word diversity were examined. Workshop notes were qualitatively reviewed to interpret the social reflection process.

## 3 Results

### 3.1 Usage Patterns and Departmental Patterns

A total of 88 participants were included in the application usage analysis during the three-month field study from October 1 to December 26, 2024. Although 56 early records were entered on September 30, after the pre-study briefing, these records were excluded from the formal usage-window summaries to ensure that the three monthly analysis windows corresponded directly to October, November, and December. During the formal intervention period, the application generated 3,480 input records containing 10,104 selected action-words. The intervention period comprised 62 nominal workdays. Participants used the application on average 39.55 ± 9.22 days (median = 40.5, range = 12-54). Each input contained 2.90 ± 0.39 selected words; 3,259 of 3,480 records (93.6%) contained all three possible word selections. All 88 participants recorded at least one entry in each of the three intervention months. The number of input records decreased over time, from 1,352 records in October to 1,246 in November, and 882 in December.

Across the full sample, participants selected all 72 available words and eight categories. Using the official English labels, the most frequently selected words were communicate (会話する; 922 selections), think deeply (思考する; 800 selections), think as a team (チームで考える; 435 selections), feel delicious (おいしくご飯を食べる; 388 selections), and appreciate (感謝する; 323 selections). "Communicate" and "think deeply" were frequently selected across departments, suggesting common workplace well-being themes related to communication and reflection. The most frequently selected categories were create (1,878, 18.6%), contact (1,771, 17.5%), hope (1,640, 16.2%), acknowledge (1,292, 12.8%), and wellness (1,256, 12.4%). These patterns suggest that the word set captured diverse aspects of workplace well-being, although some domains were more salient than others.

Beyond these common word-level tendencies, organizational differences were examined at the category level (Figure 3). The department-level heatmap provides a detailed view of the variation across the seven anonymized departments (Figure 3A); the grouped comparison highlights the broader contrast between administrative departments (Departments A-E) and research-oriented departments (Departments F-G) (Figure 3B). In the grouped view, hope-related words were more frequent in research-oriented departments than in administrative departments (22.5% vs. 12.4% of the selected words). Conversely, contact-related words (18.6% vs. 15.9%) and acknowledgment-related words (14.5% vs. 10.0%) were more frequent in administrative departments.

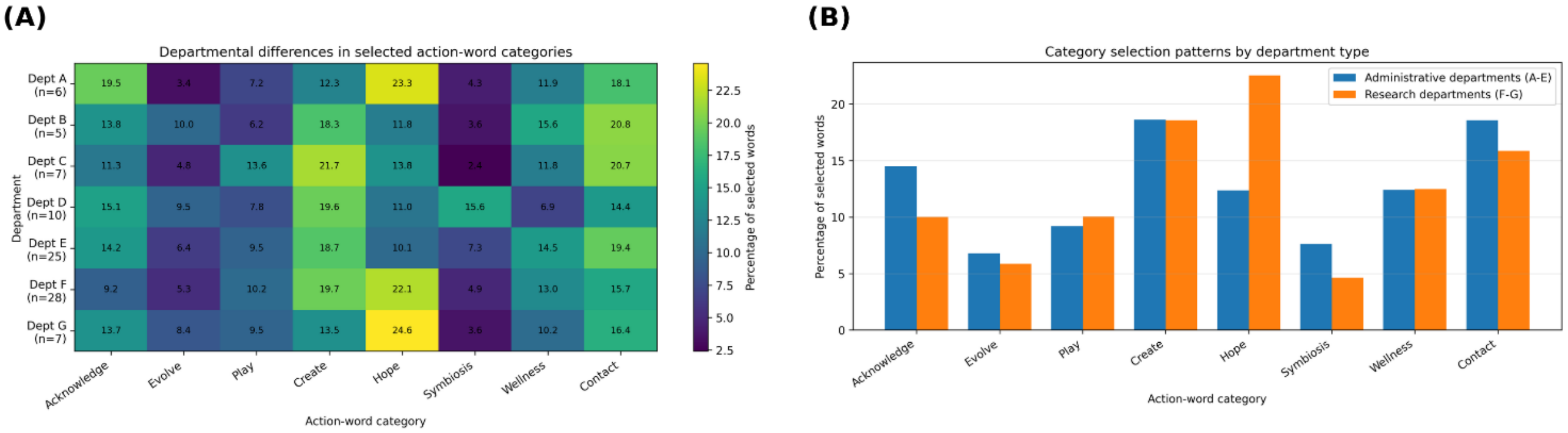

**Figure 3.** (A) Relative frequency of each category by department. (B) Category selection patterns compared between administrative departments (A-E) and research-oriented departments (F-G).

These descriptive patterns suggest that the application can visualize both broadly shared and context-specific ways in which workplace well-being-related experiences are recognized. However, these differences should be interpreted cautiously, as department sizes are uneven and the patterns may reflect participant composition, input frequency, workshop participation, and organization-specific culture as well as departmental work context.

## 3.2 Changes in PERMA Profiler Scores

PERMA Profiler scores were assessed at five time points: baseline, one month after the start, two months after the start, immediately after the intervention, and at 1-month follow-up. The overall PERMA score and five core PERMA domains were analyzed. The overall PERMA score was calculated consistently across time points using the PERMA Profiler scoring procedure.

As a sample-specific check of internal consistency, Cronbach's alpha was calculated for the internally translated Japanese PERMA items. The overall PERMA items showed high internal consistency, with $\alpha = .920$ at baseline ($n = 85$ complete item-level cases) and $\alpha = .941$ at post-intervention ($n = 88$). Domain-level alpha values ranged from .672 to .885 at baseline and from .767 to .839 at post-intervention; these domain-level estimates should be interpreted cautiously because each domain consisted of only three items.

As shown in Figure 4A, the overall PERMA score increased from $5.90 \pm 1.30$ at baseline to $6.28 \pm 1.33$ immediately after the intervention. This pre-post change was statistically supported by a paired-sample t-test (mean difference = 0.39, 95% CI [0.18, 0.60], $t[87] = 3.65$, $p < .001$, Cohen's $d_z = 0.39$). Wilcoxon signed-rank sensitivity analysis supported this conclusion ($p < .001$). In total, 61 of the 88 participants (69.3%) showed an increase in the overall PERMA score from baseline to post-intervention, 26 participants showed a decrease, and one participant showed no change.

Across the five core PERMA domains, the clearest pre-post increases were observed in positive emotion and accomplishment. Positive emotion increased from $5.67 \pm 1.53$ to $6.22 \pm 1.57$, and accomplishment increased from $5.73 \pm 1.48$ to $6.31 \pm 1.40$. Relationship and meaning also showed small increases, whereas engagement showed a modest increase that did not reach the conventional $p < .05$ threshold in the paired-sample t-test. Descriptive statistics and exploratory pre-post comparisons of the overall PERMA score and the five core domains are reported in Table 2.

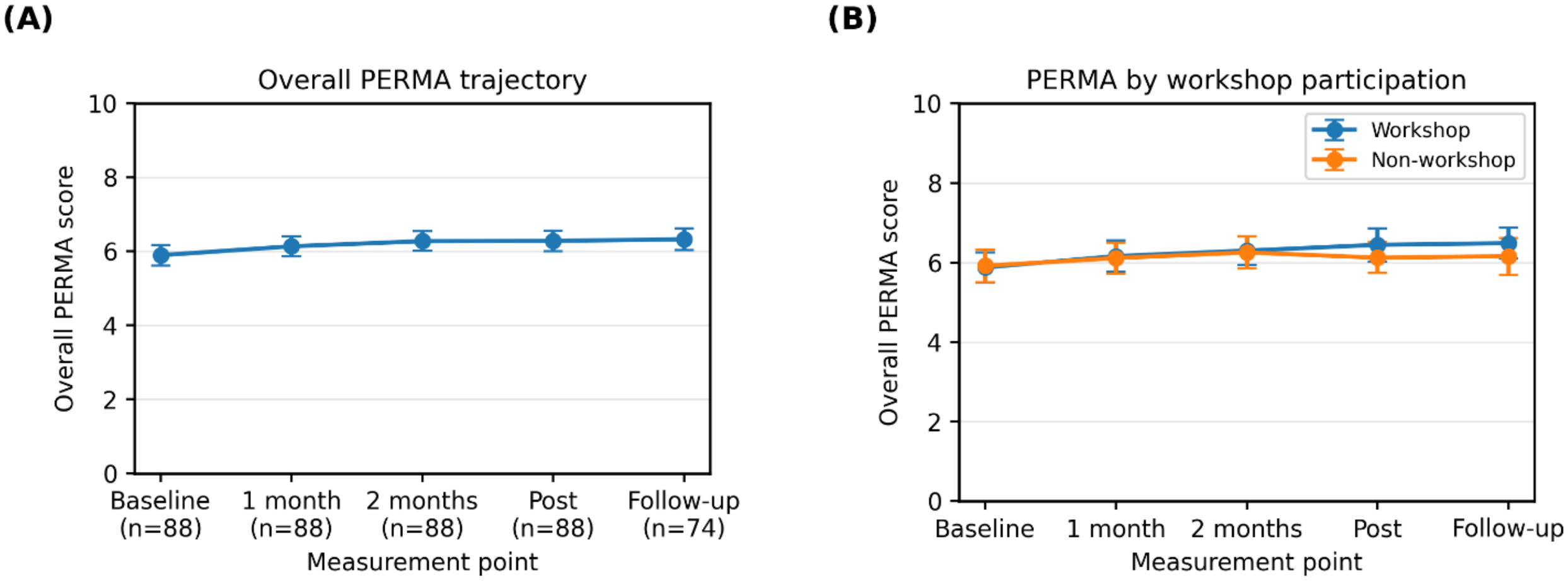

**Figure 4.** (A) Overall PERMA trajectory in the full analytic sample. (B) Overall PERMA trajectory stratified by workshop participation. Error bars represent 95% confidence intervals.

At the one-month follow-up, data were available for 74 participants. The overall PERMA score was 6.33 ± 1.28 at follow-up. Among participants with available follow-up data, the overall PERMA score remained similar from post-intervention to follow-up, changing only from 6.32 ± 1.30 to 6.33 ± 1.28, (mean difference = 0.01, 95% CI [-0.13, 0.14], t[73] = 0.11, p = .911, Cohen's dz = 0.01). Wilcoxon signed-rank sensitivity analysis also indicated no clear post-follow-up changes (p = .696). These comparisons suggest that the pre-post increase in overall PERMA did not decline during the one-month follow-up period. However, this interpretation was limited to participants with available follow-up data.

To examine the possible follow-up response bias, we compared the 74 follow-up respondents with the 14 participants without follow-up data. Follow-up respondents and nonrespondents did not obviously differ in baseline overall PERMA scores ([5.86 ± 1.31 vs. 6.09 ± 1.22], Welch's t = -0.65, p = .522, Cohen's d = -0.18), or post-intervention overall PERMA scores ([6.32 ± 1.30 vs. 6.08 ± 1.49], Welch's t = 0.55, p = .587, Cohen's d = 0.18). However, follow-up respondents had more formal-period input days than nonrespondents ([41.16 ± 8.32 vs. 31.00 ± 9.31], Welch's t = 3.81, p = .001, Cohen's d = 1.20), and selected more words overall ([119.08 ± 27.84 vs. 92.29 ± 27.53], Welch's t = 3.33, p = .004, Cohen's d = 0.96). Follow-up response rates were similar between the workshop group (38/44, 86.4%) and the non-workshop group ([36/44, 81.8%], Fisher's exact p = .772). Thus, follow-up analyses should be interpreted cautiously because the available follow-up data may overrepresent participants who engaged more frequently with the application.

**Table 2.** Mean scores for overall PERMA and the five core PERMA domains at each assessment point, with exploratory paired comparisons from baseline to post-intervention and from post-intervention to follow-up.

| Measure | Baseline | 1 month | 2 months | Post-intervention | 1-month follow-up | Baseline to post Δ | Baseline to post test | Post to follow-up Δ | Post to follow-up test |
|---|---|---|---|---|---|---|---|---|---|
| Overall PERMA | 5.90 ± 1.30 (n=88) | 6.14 ± 1.27 (n=88) | 6.28 ± 1.24 (n=88) | 6.28 ± 1.33 (n=88) | 6.33 ± 1.28 (n=74) | 0.39 | t(87) = 3.65, p < .001, dz = 0.39 | 0.01 | t(73) = 0.11, p = .911, dz = 0.01 |
| Positive emotion | 5.67 ± 1.53 (n=88) | 6.02 ± 1.55 (n=88) | 6.13 ± 1.57 (n=88) | 6.22 ± 1.57 (n=88) | 6.30 ± 1.47 (n=74) | 0.55 | t(87) = 4.07, p < .001, dz = 0.43 | 0.03 | t(73) = 0.25, p = .807, dz = 0.03 |
| Engagement | 5.86 ± 1.66 (n=88) | 6.13 ± 1.74 (n=88) | 6.19 ± 1.52 (n=88) | 6.14 ± 1.67 (n=88) | 6.29 ± 1.53 (n=74) | 0.28 | t(87) = 1.97, p = .052, dz = 0.21 | 0.08 | t(73) = 0.65, p = .516, dz = 0.08 |
| Relationships | 6.09 ± 1.49 (n=88) | 6.28 ± 1.49 (n=88) | 6.44 ± 1.46 (n=88) | 6.39 ± 1.49 (n=88) | 6.43 ± 1.42 (n=74) | 0.29 | t(87) = 2.46, p = .016, dz = 0.26 | -0.01 | t(73) = -0.13, p = .893, dz = -0.02 |
| Meaning | 5.81 ± 1.79 (n=88) | 5.96 ± 1.75 (n=88) | 6.05 ± 1.72 (n=88) | 6.14 ± 1.69 (n=88) | 6.17 ± 1.69 (n=74) | 0.33 | t(87) = 2.33, p = .022, dz = 0.25 | 0.04 | t(73) = 0.42, p = .673, dz = 0.05 |
| Accomplish-ment | 5.73 ± 1.48 (n=88) | 6.03 ± 1.44 (n=88) | 6.30 ± 1.25 (n=88) | 6.31 ± 1.40 (n=88) | 6.20 ± 1.40 (n=74) | 0.58 | t(87) = 4.09, p < .001, dz = 0.44 | -0.13 | t(73) = -1.29, p = .201, dz = -0.15 |

Note. Values are presented as means ± standard deviation. Baseline-to-post comparisons were performed for all the 88 participants. Post-follow-up comparisons included 74 participants for whom follow-up data were available. Δ indicates the mean paired change score. Wilcoxon signed-rank tests were conducted as sensitivity analyses; for the overall PERMA score, the baseline-to-post increase remained supported (p < .001), whereas the post-to-follow-up change was not (p = .696).

### 3.3 Exploratory Associations with the Sharing Workshop

The patterns of well-being sharing among workshop participants were examined by comparing the 44 participants who attended the workshop with the 44 who did not. The workshop was conducted between the one-month and two-month measurement points and functioned as a structured social reflection session. Baseline overall PERMA scores were nearly identical between the workshop group (5.88 ± 1.27) and the non-workshop group (5.92 ± 1.34; Welch's t = -0.15, p = .885, Cohen's d = -0.03).

As shown in Figure 4B, the workshop group had a numerically larger increase in the overall PERMA from baseline to post-intervention than the non-workshop group. The workshop group increased by 0.57 ± 0.93, whereas the non-workshop group increased by 0.20 ± 1.03. This group difference in pre-post change was suggestive, but not clearly supported by the conventional p < .05 threshold (Welch's t = 1.74, p = .085, and Cohen's d = 0.37). Within-group analyses showed that the workshop group increased from 5.88 ± 1.27 at baseline to 6.44 ± 1.37 at post-intervention (t[43] = 4.05, p < .001, Cohen's dz = 0.61). The non-workshop group increased from 5.92 ± 1.34 to 6.12 ± 1.28, but this within-group change was smaller and not clearly supported (t[43] = 1.32, p = .194, Cohen's dz = 0.20).

From two months to post-intervention, the workshop group maintained an upward trend, changing by 0.15 ± 0.54, whereas the non-workshop group decreased by -0.13 ± 0.73. The group difference in this change was supported by the Welch t-test (t = 2.02, p = .046, Cohen's d = 0.43), although the Mann-Whitney sensitivity analysis was less conclusive (p = .104). At the one-month follow-up, overall

PERMA scores were 6.49 ± 1.20 in the workshop group (n = 38) and 6.16 ± 1.37 in the non-workshop group (n = 36; Welch's t = 1.09, p = .278, Cohen's d = 0.26). These results suggest a more sustained upward pattern in the workshop group; however, the between-group evidence should be interpreted cautiously.

The workshop also examined the selected-word diversity (Figure 5; Table 3). In the month-based analysis, alpha diversity decreased from the second to the third month in both groups (Figure 5A). The workshop group changed from 2.64 ± 0.56 to 2.51 ± 0.51, whereas the non-workshop group changed from 2.58 ± 0.48 to 2.34 ± 0.44. The group difference in change was not clearly supported by Welch's t-test (t = 1.36, p =.178, Cohen's d = 0.29), although the decrease was numerically smaller in the workshop group. This month-based decrease should be interpreted cautiously because the number of input records and selected words also declined in the third month, which may have reduced participants' opportunities to select a broader range of words within that monthly window.

In contrast, the cumulative alpha diversity increased from the first two months to the full three-month intervention period in both groups, indicating that participants continued to accumulate a broader repertoire of selected words over time (Figure 5B). The cumulative increase was larger in the workshop group (0.11 ± 0.13) than in the non-workshop group (0.05 ± 0.09; Welch's t = 2.23, p = .028, Cohen's d = 0.48). The Mann-Whitney sensitivity analysis showed a similar tendency, but was marginal (p = .053). Thus, although the single-month diversity decreased in the third month, the cumulative diversity continued to grow, with a somewhat larger expansion among workshop participants.

In the workshop-based window analysis, alpha diversity among workshop participants changed from 2.80 ± 0.54 before the workshop to 2.75 ± 0.53 after the workshop. This mean change was not statistically significant (mean difference =-0.05, 95% CI[-0.14, 0.05], t [43] =-1.00, p =.321, Cohen's dz =-0.15). However, participants with lower pre-workshop alpha diversity tended to show larger subsequent increases, as indicated by a negative association between pre-workshop alpha diversity and pre-post change (Pearson r = -0.34, p = .025; Spearman ρ = -0.36, p = .017; Figure 5C). This suggests that the workshop may have been particularly relevant for participants whose initial selected-word repertoires were relatively narrow.

The sticky-note responses written during the workshop provided descriptive context for these exploratory patterns. Participants commonly described the application as creating an opportunity to reflect on the day, notice positive aspects of ordinary workdays, recognize their own selection tendencies, and identify what they personally experienced as positive or meaningful. Other responses referred to differences in how participants interpreted the same words, interest in others' reasons for selecting particular words, and awareness of departmental or team tendencies through aggregated visualizations. Several notes also expressed future-oriented intentions, such as trying previously unselected words, broadening one's perspective, discussing selected words with others, or using the palette as a trigger for communication.

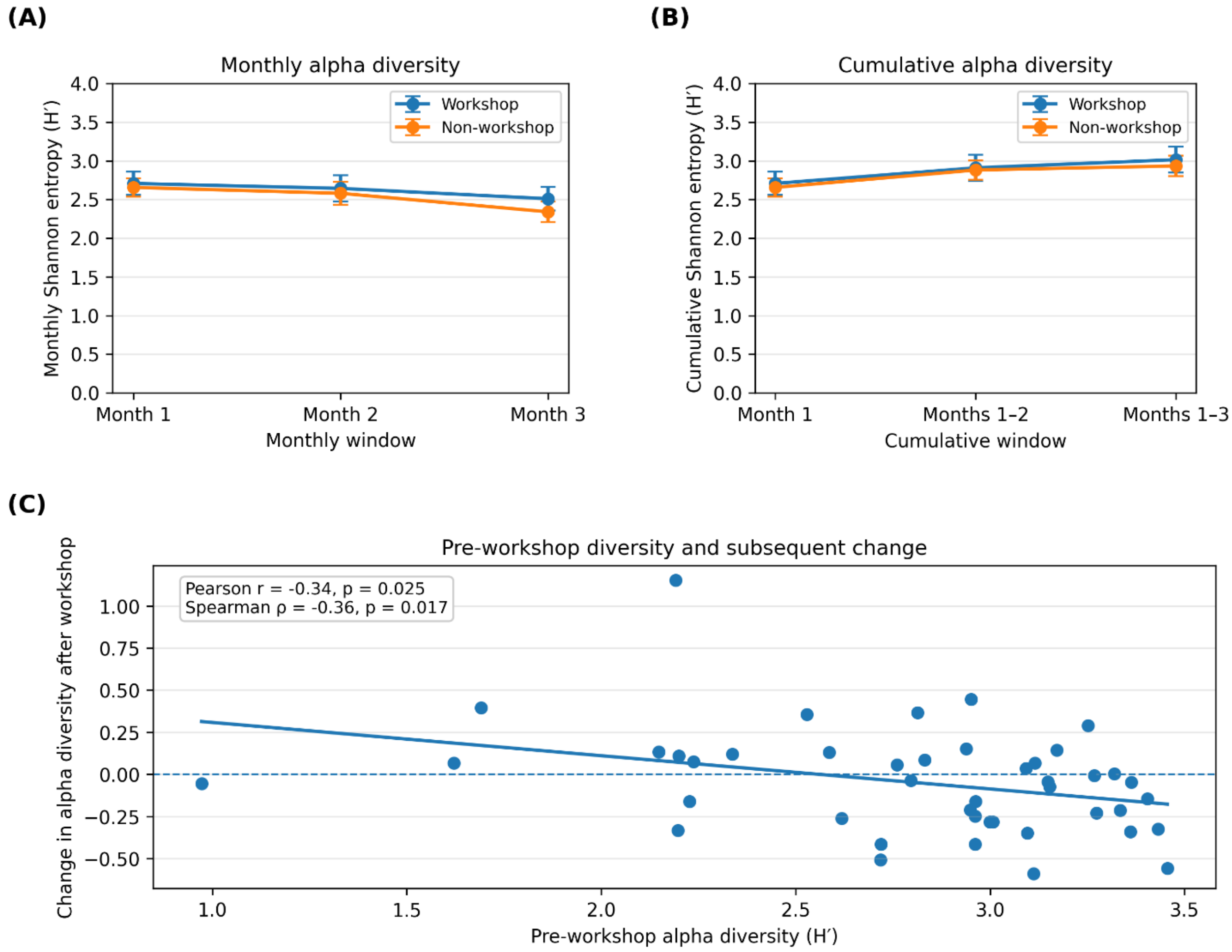

**Figure 5.** (A) Monthly alpha diversity. (B) Cumulative alpha diversity. (C) Association between pre-workshop alpha diversity and subsequent change among workshop participants. Error bars in panels A and B represent 95% confidence intervals.

**Table 3.** Shannon entropy values for monthly and cumulative selected-word-diversity windows, shown separately for workshop and non-workshop participants.

| Alpha diversity window | Workshop | Non-workshop |
|---|---|---|
| Month 1 | 2.71 ± 0.49 (n=44) | 2.66 ± 0.39 (n=44) |
| Month 2 | 2.64 ± 0.56 (n=44) | 2.58 ± 0.48 (n=44) |
| Month 3 | 2.51 ± 0.51 (n=44) | 2.34 ± 0.44 (n=44) |
| Cumulative months 1-2 | 2.91 ± 0.56 (n=44) | 2.88 ± 0.41 (n=44) |
| Cumulative months 1-3 | 3.01 ± 0.56 (n=44) | 2.93 ± 0.43 (n=44) |
| Change: Month 2 to Month 3 | -0.13 ± 0.32 | -0.24 ± 0.40 |
| Change: cumulative months 1-2 to months 1-3 | 0.11 ± 0.13 | 0.05 ± 0.09 |

Note. Values are presented as means ± standard deviation. Alpha diversity was calculated using Shannon entropy, based on the distribution of selected words within each time window.

### 3.4 Summary of Results

Overall, the results suggested three main patterns. First, the application captured 3,480 formal-period input records, and all 72 available action-words were selected, enabling the visualization of differences across departmental contexts. Second, PERMA Profiler scores increased from baseline to post-intervention and did not clearly decline at the one-month follow-up among the available cases. Third, workshop participation was associated with a numerically more sustained upward PERMA pattern and a larger cumulative increase in selected-word diversity; within the workshop group, lower pre-workshop diversity was associated with larger subsequent increases in selected-word diversity. The clearest quantitative finding was the increase in overall PERMA from baseline to post-intervention, whereas the domain-level, workshop-related, departmental, and selected-word diversity findings should be regarded as exploratory patterns that provide contextual and hypothesis-generating evidence.

## 4 Discussion

### 4.1 Summary of Main Findings

This exploratory field study examined an action-based reflection tool designed to be low-burden for workplace well-being. The observed usage patterns suggest feasibility and repeated engagement in this workplace context, although the study did not directly measure perceived burden or time required for each entry. Over the formal three-month intervention period, the participants generated 3,480 input records. They selected all 72 available action-words, suggesting that the application was used across a broad range of available vocabulary. Participants used the application on an average of 39.55 workdays, and 93.6% of the input records contained all three possible word selections. These usage patterns indicate that the daily selection task was feasible in this workplace context, although direct session duration data were unavailable.

Subjective well-being was assessed using the PERMA Profiler and increased from baseline to immediately after the intervention. Well-being did not clearly decline at the one-month follow-up among participants with available follow-up data. Workshop participants showed a numerically more sustained upward pattern in overall PERMA scores than non-workshop participants, but the between-group evidence was exploratory and should not be interpreted causally. Application log data also revealed differences in selected-word categories across departments, suggesting that the tool may help visualize how well-being-related experiences are experienced differently across work contexts.

Shannon entropy analysis provides a complementary perspective. Monthly alpha diversity decreased from the second to the third month, whereas cumulative alpha diversity increased from the first two months to the entire three-month period. This pattern suggests that short-window selected-word diversity may decline as input frequency decreases or word choices stabilize, while the cumulative range of selected words can continue to expand over time. Among the workshop participants, the mean change from the pre-workshop to the post-workshop windows in alpha diversity was not clearly supported; however, participants with lower pre-workshop alpha diversity tended to show larger subsequent increases. This finding suggests that social reflection may be particularly relevant for participants whose initial selected word repertoires are relatively narrow.

## 4.2 Designing for Low-Burden Action-Based Reflection in the Workplace

The central goal of the Well-Being Palette is to reduce the burden of daily reflection. Rather than requiring structured writing, training modules, or extended exercises, the application asks users to select up to three words after reviewing a visual palette of well-being-related action-words. This simple interaction was intended to fit the end of the workday without requiring substantial additional time or cognitive effort.

The observed usage patterns are consistent with the aim of this design. The participants generated a large number of input records during the formal intervention period, and most of these records included all three possible selections. Although the study did not record session start and end times, these log patterns suggest that the interaction was sufficiently acceptable to be repeated over time by participants in a real workplace setting. This is important because engagement and adherence are common challenges in digital well-being interventions (Baumel et al., 2019; Linardon et al., 2019; Groot et al., 2023).

An action-based format can support a particular form of reflection. By selecting words that corresponded to positive actions or experiences encountered during the day, participants not only reported their general mood state but also identified concrete aspects of daily life that they associated with well-being. Rather, the application may have provided participants with an opportunity to notice, label, and describe well-being-related experiences on their own terms. This perspective is consistent with the possibility that subjective experiences may be shaped by both external conditions and by how experiences are cognitively perceived and interpreted (Sakimura et al., 2025). Thus, the tool may function as a practical scaffold for everyday workplace reflection.

## 4.3 Interpreting Selected-Word Diversity

A key contribution of this study is the use of Shannon entropy to characterize the diversity of the selected well-being-related words. Conventional well-being assessments typically focus on well-being levels such as mean questionnaire scores. Such measures are essential, but they may not capture how broad or narrow the underlying set of well-being-related experiences is. For example, two participants may have similar PERMA scores while relying on different ranges of selected words when describing their daily positive experiences.

In this study, alpha diversity was used to quantify the distribution of the selected words within the specified time windows. This approach extends entropy-based diversity measures (widely used in ecology and other domains) to the analysis of well-being-related reflection data (Shannon, 1948; Magurran, 2004; Song et al., 2010; Yoshimura et al., 2022). However, the interpretation of this measure should be specific. Therefore, it is best understood as an index of the breadth of participants' well-being-related experiences or selected action-word repertoire.

The distinction between the monthly and cumulative alpha diversity was informative. Monthly diversity decreased from the second to the third month. This may reflect a reduced input frequency, stabilization of word choices, or a narrower set of experiences recognized within the shorter third-month window. In contrast, cumulative diversity increased from months one to two to months one to three, indicating that participants continued to accumulate a broader set of selected words across the intervention period. This contrast suggests that selected-word diversity depends on the temporal window of analysis and should be interpreted in relation to input frequency and observation period.

Critically, higher diversity should not automatically be interpreted as better well-being. A broader selected-word repertoire may indicate greater awareness of varied well-being-related experiences. However, repeatedly selecting a smaller set of personally meaningful words may also reflect stable and important sources of well-being. Thus, selected-word diversity is best viewed as a complementary lens rather than a replacement for subjective well-being measures, such as the PERMA Profiler.

Exploratory supplementary analyses further suggested that selected-word diversity and subjective well-being may capture related but distinct aspects of workplace well-being (Supplementary Figure S2). When participants were divided into high- and low-cumulative alpha diversity groups using the median Shannon entropy value, the higher-diversity group showed a descriptive tendency to maintain PERMA scores more consistently during follow-up. In contrast, the lower-diversity group showed larger intervention-period gains but a slight decline after the intervention ended. These patterns were not statistically definitive, and baseline PERMA differences were observed between groups.

Nevertheless, the findings are consistent with the possibility that selected-word diversity reflects the breadth or stability of participants' reflective well-being repertoires, complementing level-based measures such as the PERMA Profiler. As these analyses were exploratory, they should be interpreted cautiously and require replication in larger, more controlled studies.

## 4.4 Social Reflection Through the Sharing Workshop

The workshop was designed to extend individual reflections to a social setting. Participants shared their own selection patterns, explained the reasons behind their choices, listened to others' interpretations, and reviewed group-level visualizations. They also engaged in structured retrospective reflection using modified YWT prompts: what went well, what was learned, and what to try next. This process was intended to support social reflection.

The quantitative results suggest a cautious interpretation of the workshop findings. Workshop participants showed a more sustained upward pattern in overall PERMA scores, and cumulative alpha diversity increased more in the workshop group than in the non-workshop group. However, the workshops were not randomly assigned, and the evidence should be interpreted as exploratory. Moreover, the workshop-based pre-post comparison did not show a uniform mean increase in alpha diversity across all workshop participants.

The most informative pattern was the negative association between pre-workshop alpha diversity and subsequent change; participants with lower pre-workshop alpha diversity tended to show larger increases afterward. This suggests that the workshop may have been more relevant to participants whose initial selected-word repertoires were relatively narrow. Listening to how others interpreted and selected words may have helped these participants recognize additional well-being-related experiences or categories that they had not previously considered.

The descriptive workshop-note patterns reported in the Results are consistent with this interpretation. Recurrent themes include daily noticing, recognition of diversity in others' perspectives, self-understanding through one's own selection patterns, visualization of organizational tendencies, future-oriented expansion of word choices and possible actions, and palette-triggered communication. These observations were not treated as primary outcomes, but they helped to contextualize how the workshop may have supported reflection. This interpretation is consistent with social and experiential learning perspectives, in which exposure to others' experiences can expand how individuals interpret and act within their environment (Bandura, 1977; Kolb, 1984).

### 4.5 Organizational Visualization and Practical Implications

The application also provides a way to visualize differences in selected well-being-related words across departments. Across the full sample, communicate and think deeply were frequently selected, suggesting common patterns in how the participants recognized positive workplace experiences. At the same time, research-oriented departments showed a relatively higher selection of hope-related words. In contrast, administrative departments showed a relatively higher selection of contact- and acknowledgment-related words.

These departmental patterns should be interpreted with caution. They may reflect work content, departmental culture, participant composition, input frequency, workshop participation, or company-specific contexts. Nevertheless, the ability to visualize such patterns may have practical value. Organizations often rely on aggregate well-being scores, which can obscure the differences in the recognition and expression of well-being across groups. Selected-word visualizations may provide a concrete basis for dialogue about how well-being-related experiences are recognized across work contexts.

Practically, the proposed approach may be useful in two ways. First, as an individual reflection tool, it offers a way to notice and record daily well-being-related experiences. Second, when combined with visualizations and workshops, it may support team-level dialogue on how well-being is experienced in specific work contexts. These uses should be understood as exploratory applications rather than as established intervention effects.

### 4.6 Limitations and Future Directions

This study had several limitations. First, it was conducted as a field experiment in a real workplace and did not include randomized control conditions. Therefore, the observed changes in PERMA scores, selected-word diversity, and workshop-related patterns should be interpreted as exploratory associations, rather than causal effects of the application or workshop. In addition, multiple exploratory analyses were conducted without formal correction for multiple comparisons; therefore, nominally significant findings outside the main baseline-to-post-intervention overall PERMA comparison should be interpreted cautiously as hypothesis-generating. Future studies should include control or comparison conditions such as alternative reflection tools, blank logging systems, or delayed-intervention designs.

Second, the application log data were based on participants' self-recognition and recall of positive experiences at the end of each workday. Therefore, increases in alpha diversity should be interpreted as reflecting the diversification of selected reflective words or recognized well-being-related experiences. It is not definitive evidence that participants' actual behaviors became more diverse. Future studies should combine word-selection data with ecological momentary assessments, qualitative interviews, passive additional indicators, or third-party observations, where appropriate. In addition, Shannon entropy is sensitive to the number of observations within a time window. Although entropy was calculated from relative frequencies, monthly estimates may still have been influenced by differences in input days or total selected words across participants and months. Future studies should consider sensitivity analyses such as normalized entropy, rarefaction-based comparisons, or model-based adjustment for input frequency.

Third, the follow-up analyses were based on available cases. Although the follow-up respondents and nonrespondents did not clearly differ in baseline or post-intervention overall PERMA scores, the follow-up respondents had more application input days and selected more words during the intervention period. Thus, the follow-up findings may overrepresent participants who engaged more frequently with the application.

Fourth, the sample comprised only 88 participants from a single private-sector corporate research institute. Although the participants came from multiple departments, the findings may reflect organization-specific culture, policies, or work practices. Word sets and category labels were also developed in the context of Japanese corporate workplaces. However, its applicability to other cultures, languages, industries, and occupational environments is yet to be examined.

Fifth, the word set used in the exploratory field study was not independently validated as a psychometric instrument. Therefore, the Well-Being Palette vocabulary should be understood as a configurable reflection vocabulary, rather than a validated scale. Future research should examine how different word sets, translations, and cultural adaptations impact both user experience and analytic outcomes.

Sixth, the Japanese version of the PERMA Profiler used in this study was internally translated for this exploratory field study. Although the original item structure, domain assignments, and response format were retained and the wording was reviewed internally, formal back-translation and independent psychometric validation were not conducted. Therefore, measurement equivalence with the original English version cannot be fully assumed.

Seventh, the sharing workshop was a multi-component social reflection intervention. The present study could not isolate which components (individual disclosure, peer questioning, group-level visualization, or structured reflection) contributed the most strongly to the observed patterns. In addition, workshop participation was not randomly assigned. Differences between workshop and non-workshop participants may reflect assignment procedures, departmental composition, or other baseline characteristics, despite participants being assigned based on demographic balance rather than prior application-use frequency.

Finally, the definition of alpha diversity requires further theoretical and empirical refinement. A higher selected-word diversity may indicate a broader reflective repertoire, whereas a lower diversity may reflect stable and personally meaningful sources of well-being. Future studies should examine how selected-word diversity relates to subjective well-being, engagement, job characteristics, and actual behaviors over longer periods.

### 4.7 Conclusion

This study proposes and evaluates the Well-Being Palette, an action-based reflection tool designed to be low-burden for workplace well-being. Over a three-month exploratory field study, participants repeatedly used the application to select well-being-related action-words, and the overall PERMA scores increased from baseline to post-intervention. Shannon entropy provided a complementary way to characterize the diversity of the selected reflective words, revealing patterns that were not captured by mean well-being scores alone. The sharing workshop further suggested that social reflection may be particularly relevant for participants with initially narrow selected-word repertoires.

Overall, the findings suggest that brief action-word reflection can be integrated into workplace routines and may support the recognition of well-being-related experiences. At the same time, the results should be interpreted cautiously because the study was exploratory and non-randomized. Further controlled and cross-contextual research is required to evaluate whether this approach can reliably support workplace well-being and broaden how workers recognize, describe, and share their well-being-related experiences.

## 5 Conflict of Interest

This work was supported by Toyota Central R&D Laboratories, Inc. (Japan). NM, TN, TT, KI, and KK are employees of Toyota Central R&D Laboratories, Inc. Patent applications have been filed for the technology described in this study. NM, TN, TT, KI, and KK are the inventors of these patents.

## 6 Author Contributions

NM: Conceptualization, Methodology, Resources, Supervision, Validation, Writing - original draft, Writing - review & editing. TN: Conceptualization, Software, Data curation, Formal analysis, Methodology, Investigation, Visualization, Writing - original draft, Writing - review & editing. TT: Conceptualization, Data curation, Formal analysis, Methodology, Visualization, Writing - review & editing. KI: Conceptualization, Data curation, Formal analysis, Methodology, Writing - review & editing. KK: Conceptualization, Methodology, Project administration, Writing - review & editing.

## 7 Funding Statement

The author(s) declare that they received financial support for the research and the publication of this article. Funding was provided by Toyota Central R&D Laboratories, Inc. Funding did not influence any aspect of the study design or report of results.

## 8 Ethical Considerations

This study involving human participants was reviewed and approved by the Toyota Central R&D Laboratories, Inc. Research Ethics Review Committee (approval no. 24A-22). All participants provided informed consent before participation. Analyses were conducted using anonymized participant identifiers.

## 9 Acknowledgments

We would like to express our gratitude to all the participants who took part in this field study over the course of three months. We also thank Hirotaka Kaji, Masahiro Nishio, Heishiro Toyoda, and Yuhei Yamaguchi of Toyota Motor Corporation, as well as members of the Emotional Well-Being Study Group, for participating in discussions on the selection of words for the Well-Being Palette and broader activities around using it. We thank Editage (www.editage.jp) for English language editing.

## 11 Supplementary Material

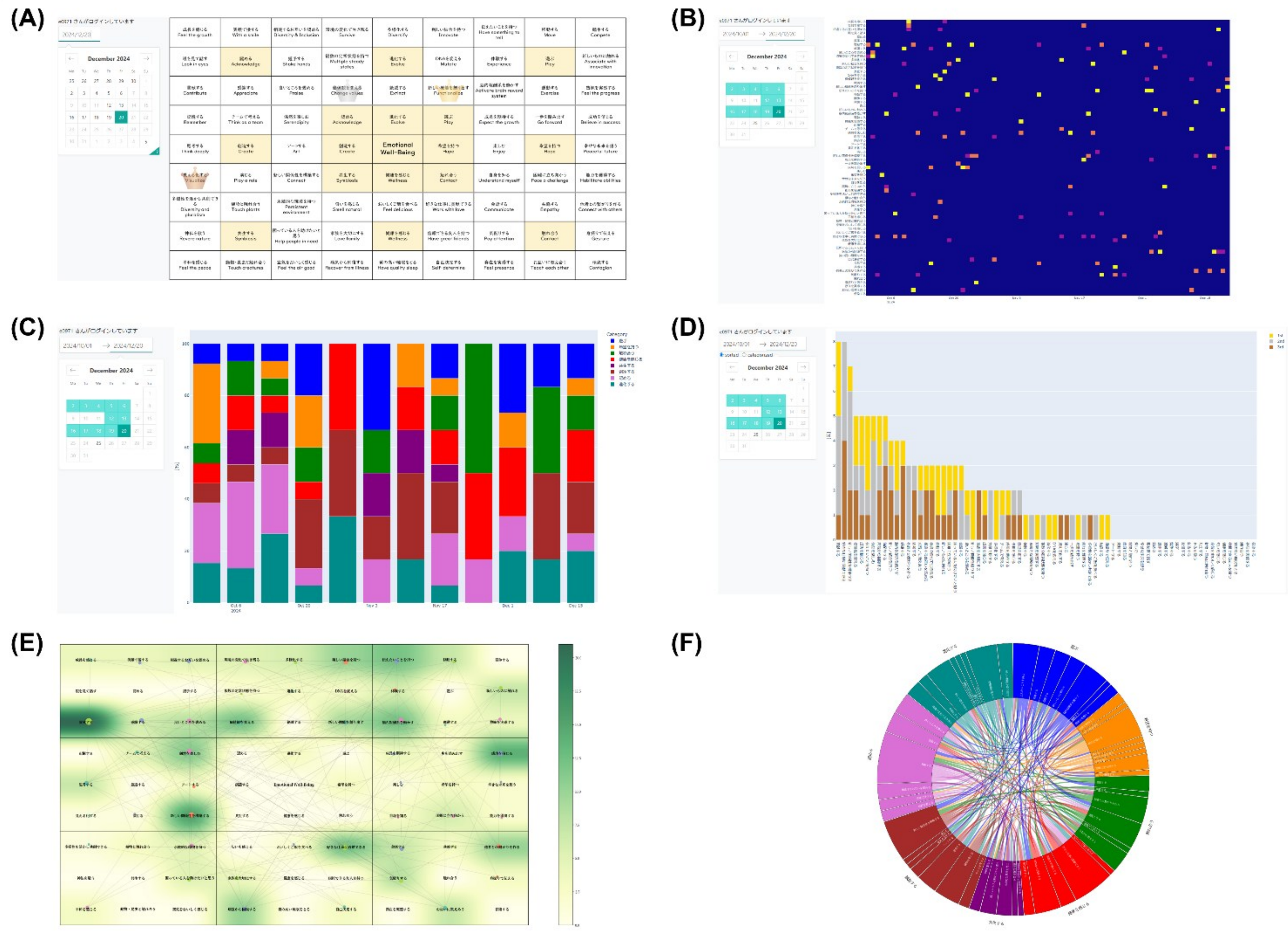

**Supplementary Figure S1.** (A) Selected-day review. (B) Word-by-date heatmap. (C) Weekly category composition. (D) Selected-word frequency ranking by selection order. (E) Palette-based frequency and co-occurrence map. (F) Circular co-occurrence visualization.

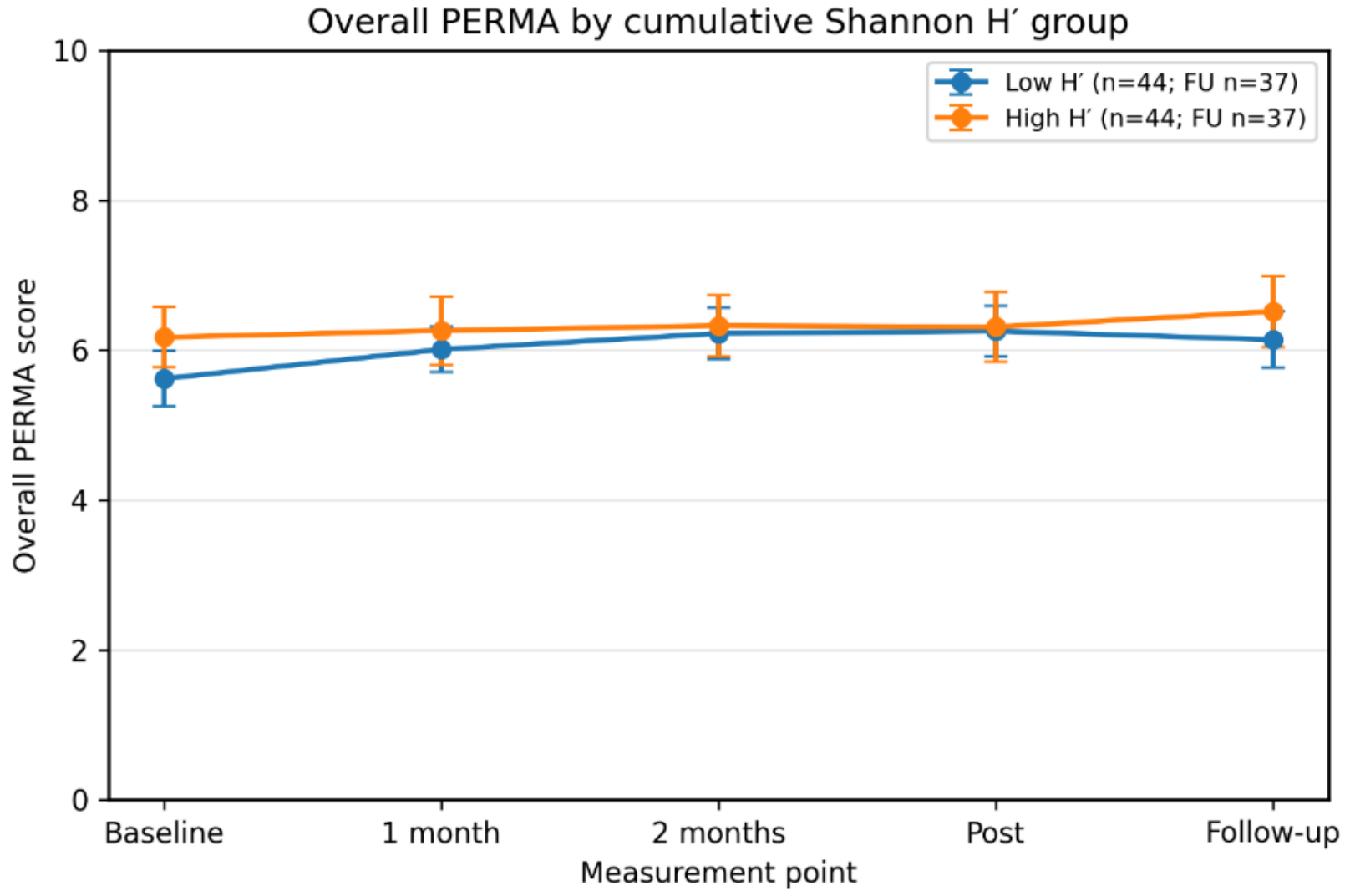

**Supplementary Figure S2.** Exploratory PERMA trajectories for participants with higher and lower cumulative selected-word diversity, defined using a median split of Shannon entropy calculated across the full three-month intervention period.

**Supplementary Table S1.** Full set of well-being-related action words used in the application, including Japanese labels and official English labels for categories and words. The categories Acknowledge, Play, Hope, Wellness, and Contact were informed by concepts discussed in previous well-being research. The Create category represents creative and generative actions associated with worker well-being, while the Evolve category was derived from values considered important in organizational culture. The Symbiosis category was included to represent harmonious coexistence with nature, a concept associated with spiritual well-being in some cultures.

| Category (jp) | Category (en) | Word (jp) | Word (en) |
|---|---|---|---|
| 認める | Acknowledge | 成長を感じる | Feel the growth |
| 認める | Acknowledge | 笑顔で接する | With a smile |
| 認める | Acknowledge | 相違するお互いを認める | Diversity & Inclusion |
| 認める | Acknowledge | 眼を見て話す | Look in eyes |
| 認める | Acknowledge | 認める | Acknowledge |
| 認める | Acknowledge | 握手する | Shake hands |
| 認める | Acknowledge | 貢献する | Contribute |
| 認める | Acknowledge | 感謝する | Appreciate |
| 認める | Acknowledge | 良いところを褒める | Praise |
| 進化する | Evolve | 環境の変化で生き残る | Survive |
| 進化する | Evolve | 多様化する | Diversify |
| 進化する | Evolve | 新しい結合を持つ | Innovate |
| 進化する | Evolve | 複数の定常状態を持つ | Multiple steady states |
| 進化する | Evolve | 進化する | Evolve |
| 進化する | Evolve | DNA を変える | Mutate |
| 進化する | Evolve | 価値観を変える | Change values |
| 進化する | Evolve | 絶滅する | Extinct |
| 進化する | Evolve | 新しい機能を創り出す | Functionalize |
| 遊ぶ | Play | 伝えたいことを持つ | Have something to tell |

| Category (jp) | Category (en) | Word (jp) | Word (en) |
|---|---|---|---|
| 遊ぶ | Play | 移動する | Move |
| 遊ぶ | Play | 競争する | Compete |
| 遊ぶ | Play | 体験する | Experience |
| 遊ぶ | Play | 遊ぶ | Play |
| 遊ぶ | Play | 新しいものに触れる | Associate with innovation |
| 遊ぶ | Play | 脳内報酬系を働かす | Activate brain reward system |
| 遊ぶ | Play | 運動する | Exercise |
| 遊ぶ | Play | 熟練を実感する | Feel the progress |
| 創造する | Create | 記憶する | Remember |
| 創造する | Create | チームで考える | Think as a team |
| 創造する | Create | 偶然を楽しむ | Serendipity |
| 創造する | Create | 思考する | Think deeply |
| 創造する | Create | 創造する | Create |
| 創造する | Create | アートする | Art |
| 創造する | Create | 見える化する | Visualize |
| 創造する | Create | 演じる | Play a role |
| 創造する | Create | 新しい関係性を構築する | Connect |
| 希望を持つ | Hope | 成長を期待する | Expect the growth |
| 希望を持つ | Hope | 一歩を踏み出す | Go forward |
| 希望を持つ | Hope | 成功を信じる | Believe in success |
| 希望を持つ | Hope | 楽しむ | Enjoy |
| 希望を持つ | Hope | 希望を持つ | Hope |
| 希望を持つ | Hope | 幸せな未来を想う | Peaceful future |
| 希望を持つ | Hope | 自身を知る | Understand myself |
| 希望を持つ | Hope | 困難に立ち向かう | Face a challenge |
| 希望を持つ | Hope | 能力を獲得する | Habilitate abilities |
| 共生する | Symbiosis | 多様性を活かし共創できる | Diversity and pluralism |
| 共生する | Symbiosis | 植物と触れ合う | Touch plants |
| 共生する | Symbiosis | 永続的な環境を持つ | Persistent environment |
| 共生する | Symbiosis | 神仏を敬う | Revere nature |
| 共生する | Symbiosis | 共生する | Symbiosis |
| 共生する | Symbiosis | 困っている人を助けたいと思う | Help people in need |
| 共生する | Symbiosis | 平和を感じる | Feel the peace |
| 共生する | Symbiosis | 動物・昆虫と触れ合う | Touch creatures |
| 共生する | Symbiosis | 空気をおいしく感じる | Feel the air good |
| 健康を感じる | Wellness | 匂いを感じる | Smell natural |
| 健康を感じる | Wellness | おいしくご飯を食べる | Feel delicious |
| 健康を感じる | Wellness | 好きな仕事に貢献できる | Work with love |
| 健康を感じる | Wellness | 家族を大切にする | Love family |
| 健康を感じる | Wellness | 健康を感じる | Wellness |
| 健康を感じる | Wellness | 信頼できる友人を持つ | Have great friends |
| 健康を感じる | Wellness | 病気から回復する | Recover from illness |
| 健康を感じる | Wellness | 質の高い睡眠をとる | Have quality sleep |
| 健康を感じる | Wellness | 自己決定する | Self-determine |
| 触れ合う | Contact | 会話する | Communicate |
| 触れ合う | Contact | 共感する | Empathy |
| 触れ合う | Contact | 他者との繋がりを作る | Connect with others |
| 触れ合う | Contact | 気配りする | Pay attention |

| Category (jp) | Category (en) | Word (jp) | Word (en) |
| --- | --- | --- | --- |
| 触れ合う | Contact | 触れ合う | Contact |
| 触れ合う | Contact | 身振りで伝える | Gesture |
| 触れ合う | Contact | 存在を実感する | Feel presence |
| 触れ合う | Contact | お互いに教えあう | Teach each other |
| 触れ合う | Contact | 伝染する | Contagion |

**Supplementary Table S2.** Internal Japanese translation of the 23-item PERMA Profiler used in this study, including item codes, Japanese item wording, and 0-10 response anchors.

| Label | Question | Response Anchor(= 0) | Response Anchor(= 10) |
| --- | --- | --- | --- |
| A1 | あなたが自分自身の目標達成に向けて進んでいると感じる時間は、どのくらいありますか。 | 全くない | いつもある |
| E1 | あなたは自分がしていることに夢中になることが、どのくらいありますか。 | 全くない | いつもある |
| P1 | 通常、あなたはどのくらいの頻度で喜びを感じますか。 | 全く感じない | いつも感じる |
| N1 | 通常、あなたはどのくらいの頻度で不安を感じますか。 | 全く感じない | いつも感じる |
| A2 | あなたは自分で設定した重要な目標をどのくらいの頻度で達成できますか。 | 全くできない | いつもできる |
| H1 | 通常、あなたの健康状態はどうですか。 | 非常に悪い | 非常に良い |
| M1 | 通常、あなたはどの程度目的や意味のある生活を送っていますか。 | 全く送っていない | 完全に送っている |
| R1 | あなたが他者からの助けや支援を必要とするとき、それをどの程度受けていますか。 | 全く受けていない | 完全に受けている |
| M2 | 通常、あなたは自分のしていることが、どの程度価値があり、やりがいがあると感じますか。 | 全く感じない | とても強く感じる |
| E2 | 通常、あなたは物事にどの程度わくわくしたり、興味を感じたりしますか。 | 全く感じない | とても強く感じる |
| Lon | あなたは日常生活でどのくらい孤独を感じますか。 | 全く感じない | とても強く感じる |
| H2 | あなたは現在の健康状態にどのくらい満足していますか。 | 全く満足していない | 完全に満足している |
| P2 | 通常、あなたはどのくらいの頻度で前向きな気持ちを感じますか。 | 全く感じない | いつも感じる |
| N2 | 通常、あなたはどのくらいの頻度で怒りを感じますか。 | 全く感じない | いつも感じる |
| A3 | あなたはどのくらいの頻度で、責任を果たせますか。 | 全く果たせない | いつも果たせる |
| N3 | 通常、あなたはどのくらいの頻度で悲しいと感じますか。 | 全く感じない | いつも感じる |
| E3 | 楽しいことをしている時、時が経つのを忘れることがどのくらいありますか。 | 全くない | いつもある |
| H3 | 同年齢で同性の人と比べて、あなたの健康状態はどうですか。 | 非常に悪い | 非常に良い |
| R2 | あなたはどの程度自分が愛されていると感じますか。 | 全く感じない | とても強く感じる |
| M3 | あなたの人生にはどの程度方向性があると感じますか。 | 全く感じない | とても強く感じる |
| R3 | あなたは自分自身の人間関係に、どのくらい満足していますか。 | 全く満足していない | 完全に満足している |
| P3 | 通常、あなたはどの程度満足していると感じますか。 | 全く感じない | とても強く感じる |
| Hap | すべてのことを考え合わせて、あなたは自分がどのくらい幸せだと思いますか。 | 全く幸せと思わない | 完全に幸せと思う |